\documentclass[pre,superscriptaddress,twocolumn,amsmath,amssymb,aps]{revtex4}
\usepackage{graphicx,subfig} 
\usepackage[utf8]{inputenc}

\renewcommand{\phi}{\varphi}

\begin{document} 
\title{Structure and dynamics of topological defects in a glassy liquid on a negatively curved manifold}

\author{François Sausset}
\email{sausset@lptmc.jussieu.fr}
\affiliation{Institut de Physique Théorique, CEA, CNRS URA 2306, F-91191 Gif sur Yvette, France}

\author{Gilles Tarjus}
\email{tarjus@lptmc.jussieu.fr}
\affiliation{LPTMC, CNRS-UMR7600, Université Pierre et Marie Curie, boîte 121, 4 Place Jussieu, 75252 Paris cedex 05, France} 

\author{David R. Nelson}
\email{nelson@physics.harvard.edu}
\affiliation{Department of Physics, Harvard University, Cambridge, Massachusetts 02138, USA}

\date{\today}
\begin{abstract}
We study the low-temperature regime of an atomic liquid on the hyperbolic plane by means of molecular dynamics simulation and we compare the results to a continuum theory of defects in a negatively curved hexagonal background. In agreement with the theory and previous results on positively curved (spherical) surfaces, we find that the atomic configurations consist of isolated defect structures, dubbed ``grain boundary scars'', that form around an irreducible density of curvature-induced disclinations in an otherwise hexagonal background. We investigate the structure and the dynamics of these grain boundary scars.
\end{abstract}

\maketitle

\section{Introduction}

More than a century after Thompson's considerations about the organization of repulsive charged particles in spherical shells~\cite{thomson04}, the problem of ordering and minimum-energy configurations of particles on curved surfaces has attracted renewed interest~\cite{moore97,bowick00,nelson02,bowick09}. A theory has recently been proposed which derives ground-state particle configurations on 2-dimensional (2D) curved manifolds from the effective free-energy of relevant interacting topological defects, disclinations and dislocations~\cite{bowick00,bowick02,travesset05,bowick06,bowick07}. It has been mainly applied to ``spherical crystals'', \textit{i.e.} defected triangular tilings on the surface of a sphere, for which it predicts the appearance of stable ``grain boundary scars'' formed by strings of dislocations emanating from an irreducible curvature-induced disclination and terminating in the crystal. Such defect structures are energically forbidden in flat (Euclidean) space and result from the nonzero curvature of space. This prediction, as well as more specific ones concerning either the structure or the dynamics of the defects, have been confirmed in both simulations~\cite{bowick09} and experiments~\cite{bausch03,einert05,lipowsky05} on spherical crystals.

The purpose of the present work is to test the theoretical predictions on 2D manifolds of constant \emph{negative} curvature. Negative curvature brings in new features compared to the more familiar positive curvature of spherical geometry. First, a homogeneous space of constant negative curvature can be of infinite extent: in 2D, this is the hyperbolic plane $H^2$. As a result, a thermodynamic limit and \textit{bona fide} phase transitions may be envisaged, contrary to what occurs on a sphere. Secondly, the hyperbolic plane as a whole cannot be embedded in 3D Euclidean space so that physical realizations of negatively curved manifolds in soft and hard condensed matter do not have constant curvature. They can be negatively curved locally, in the form of a saddle point, or in a more extended way, as on the inner wall of a torus~\cite{bowick09}. They can also be of arbitrary spatial extent with a negative Gaussian curvature only on average, such as the ``minimal surfaces'' found for example in bicontinuous phases of amphiphilic bilayers~\cite{sadoc89}.

Here, we consider an equilibrium system of atoms interacting with a circularly symmetric pair potential on the hyperbolic plane, which we investigate through molecular dynamics simulation. This model has been introduced and already studied in the frustration-based approach of glasses~\cite{nelson83,rubinstein83,nelson02}, of supercooled liquids and the glass transition~\cite{tarjus05,sausset08,sausset09}, and of jamming~\cite{modes07,modes08}. In the present context of low-energy configurations of particles and associated topological defects, we focus on the model at low temperature, in a regime where the spatial extension of the local hexagonal/hexatic order characteristic of the liquid has saturated due to the frustrating effect of space curvature~\cite{sausset09}. We find that the atomic configurations consist of rare defect structures in an otherwise 6-fold coordinated background. At the temperatures at which equilibration is still possible on the simulation time scale, the system is disordered at long length scales and appears as a very slowly relaxing ``glassy'' liquid. Interestingly, the defect structures are organized much like what has been found in spherical geometry and predicted by the continuum theory of defects~\cite{bowick00,bowick02,travesset05,bowick06,bowick07}. They form grain boundary scars emanating from an irreducible number of disclinations forced in the medium by the topology of the embedding manifold. By changing the magnitude of the curvature, we also confirm, at least in a semi-quantitative way, several predictions of the theory.

The rest of the article is organized as follows. In section II, we introduce the atomistic model and we shortly describe basic notions and specific features coming with the hyperbolic geometry as well as the way we handle them in the computer simulations. In section III, we review the continuum theory of defects and apply it to negatively curved 2D manifolds. Section IV is devoted to the simulation results and their comparison with theoretical predictions. Finally, we give some concluding remarks in section V.

\section{Model, simulation, and basic notions}
\label{sec:model}

We consider a one-component atomic liquid embedded in the hyperbolic plane $H^2$. The latter is a homogeneous surface of constant negative curvature $K=-\kappa^2$, the associated hyperbolic metric being obtained from $\mathrm{d}s^2 =\mathrm{d}r^2  + \kappa^{-2} \sinh^2(\kappa r) \, \mathrm{d}\phi^2 $ with $(r,\phi)$ the appropriate polar coordinates~\cite{hilbert,coxeter}. The atoms interact through the Lennard-Jones potential
\begin{equation}
\label{eq_LJ}
v(r) = 4 \epsilon \left[\left(\frac{\sigma}{r} \right)^{12} - \left(\frac{\sigma}{r} \right)^{6}\right] 
\end{equation}
where $r$ is the geodesic distance between atom centers. We have investigated this model by means of molecular dynamics simulation. The simulations have been performed at fixed density $\rho \sigma^2 \simeq 0.85$ in the NVE ensemble, and several curvature magnitudes, parametrized by $\kappa \sigma$, have been considered. We have restricted our study to ``small'' curvatures ($\kappa \sigma \leq 0.2$) for which the local arrangement of the atoms is the same as in flat (Euclidean) space, \textit{i.e.} hexagonal. Due to the curvature-induced frustration, hexagonal tiling of the whole plane is however forbidden and an irreducible density of defects must be present. For larger curvatures ($\kappa \sigma \gtrsim 1$), local heptagonal, octagonal, etc..., atomic arrangements become preferred in the liquid and, for specific ranges of values, this may lead to unfrustrated (quasi) long-range ordering in regular lattices with coordination numbers $z\geq 7$ at low temperature~\cite{rubinstein83}.

Hyperbolic geometry requires extending the usual methods and algorithms used in Euclidean space in order to account for boundary effects (which are never negligible compared to bulk contributions), parallel transport of vectors (which is not globally defined on curved manifolds), and visualization (since $H^2$ cannot be embedded in $3D$ Euclidean space). Details on the procedures are given elsewhere~\cite{sausset08,sausset09}. We just mention here that, to study bulk behavior, we have implemented periodic boundary conditions~\cite{sausset07}. Quite generally, periodic boundary conditions can be envisaged from both a ``geometrical'' and a ``topological'' perspective. From a geometrical point of view, a periodic boundary condition amounts to replicating a primitive cell containing the system of interest in order to generate a tiling of the whole space. From a topological one, a periodic boundary condition implies providing a rule to pair the edges of the primitive cell, which leads, after identification of the paired edges, to a ``quotient space'' that can be represented as a multi-connected compact manifold. The primitive cell together with the edge pairing rule define a ``fundamental polygon''~\cite{sausset07}. On the hyperbolic plane, there is an infinite number of possibilities. A given fundamental polygon is then characterized by an area, or equivalently by the genus of the associated compact manifold (see below Eq.~\eqref{eq_gauss_bonnet2}), a number of sides, which for regular polygons and for a given genus fixes the associated tiling, and a pairing of its sides. In the present study, we have used two different boundary conditions. The first one corresponds to an octagonal primitive cell whose quotient space can be topologically described as a compact manifold of genus $g=2$ (a ``2-hole torus''); the associated tiling of the hyperbolic plane $H^2$ is denoted by $\{8,8\}$, meaning that $8$ octagons meet at each vertex (more generally, the Schläfli notation $\{p,q\}$ denotes a tiling in which $q$ regular polygons with $p$ edges meet at each vertex). The other periodic boundary condition is based on a a polygonal cell with $14$ edges and a quotient space with genus $g=3$ (``a 3-hole torus''); the associated tiling of $H^2$ is a $\{14,7\}$ one. Choosing these two different conditions allows us to check that the observed behavior of the liquid does correspond to bulk behavior with negligible finite-size effect: the area of the $14$-gon is indeed twice bigger than the area of the octagon whereas, as will be shown below, the measured quantities are essentially identical in both cases.

Visualization, on the other hand, is obtained through the Poincaré disk model~\cite{hilbert,coxeter}. This is a projection of the whole hyperbolic plane $H^2$ onto a unit disk. The projection is conformal (angles are preserved) but not isometric (distances are deformed and shrink as one reaches the disk boundary). Others representations of $H^2$ exist, each having its own advantages: the Poincaré half-plane, the Klein model, the hyperboloid of revolution endowed with the Lorentz metric ($\mathrm{d}s^2 = \mathrm{d}x^2 + \mathrm{d}y^2 - \mathrm{d}z^2$)~\cite{hilbert,coxeter}. An often used way to represent $H^2$ in real space is the so-called pseudosphere: it does not need any projection, has a constant negative curvature, and is of infinite extent. Its shape is defined by the surface of revolution of a tractrix~\cite{coxeter} around its asymptote and can be roughly seen as a trumpet horn of infinite height. We have however chosen the Poincaré disk model because it is well suited for the visualization of particles and defects.

\begin{figure*}
	\begin{center}
		\subfloat[]{
		\includegraphics[draft=false,width=8cm]{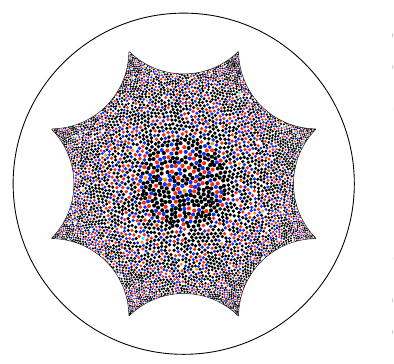}}
		\hspace{1cm}
		\subfloat[]{
		\includegraphics[draft=false,width=8cm]{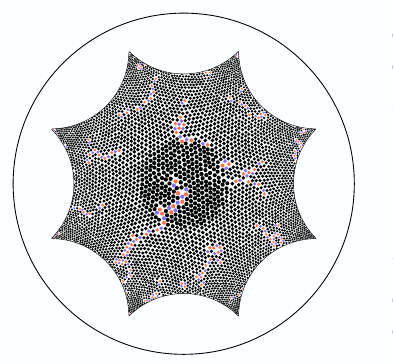}} 
	\end{center}
	\caption{ \label{fig:atomicConfs} (Color online) Atomic configurations in the Poincaré disk representation of $H^2$. Two temperatures above and below the ordering temperature $T^*\simeq 0.75$ in flat space are shown: (a) $T/T^* \simeq 2.4$, (b) $T/T^* \simeq 0.52$. From the Delaunay triangulation, atoms are colored according to their coordination: black for $6$ (hexagonal), red for $7$ (negative disclination of charge $-\pi/3$) and blue for $5$ (positive disclination of charge $+\pi/3$); other colors correspond to disclinations of higher charge magnitude. Here, $\kappa \sigma = 0.05$ and the octagonal periodic boundary condition (pbc) is used.} 
\end{figure*}

In 2D space, topological defects are point-like. A microscopic definition of the elementary defects, which are disclinations in a reference hexagonal crystal, is possible through a Delaunay triangulation of the atomic configurations. The procedure can be generalized to the hyperbolic plane, for which it has been shown to be unique when the ensemble of points formed by the atomic centers is dense enough~\cite{leibon00}, a condition always satisfied in our system. From the Delaunay construction, one can then determine the coordination number of each atom in the liquid. When $\kappa \sigma \leq 0.2$ and at low temperature $T$, most atoms are $6$-fold coordinated, which corresponds to a local hexagonal arrangement, and defects are located on atoms whose coordination number differs from $6$: for instance, a $7$-fold coordinated atom corresponds to a negative disclination of topological charge $-\pi/3$ whereas a $5$-fold one corresponds to a positive disclination of charge $\pi/3$; disclinations of higher charges $\pm q \pi/3$ with an integer $q\geq 2$ are very rare at low $T$. For illustration, two configurations at high $T$ (above the ordering transition $T^*$ in flat space) and low $T$ (significantly below $T^*$) are shown in Fig.~\ref{fig:atomicConfs}. In this work, we only focus on defect structures at low temperature, when the spatial extension of the local hexagonal order (which grows as $T$ decreases) has saturated as a result of curvature-induced frustration~\cite{sausset09}.

In 2D space, there are two important identities, the Euler-Poincar\'e relation and the Gauss-Bonnet theorem~\cite{hilbert,coxeter}. The former only involves topology and for $N$ points on a compact surface that is triangulated via the Delaunay construction, it reads
\begin{equation}
\label{eq_euler_poincare}
\frac{N}{6}\left(6-c \right) = \chi = -2\left(g-1 \right),
\end{equation}
where $c$ is the average coordination number of the points and where we have used the definition of the Euler characteristic of the surface in terms of the genus, $\chi = -2(g-1)$. On the other hand, for a surface $\Sigma$ of total area $A$ embedded in the hyperbolic plane $H^2$ of curvature $K=-\kappa^2$, the Gauss-Bonnet theorem states that
\begin{equation}
\label{eq_gauss_bonnet}
\int_{\Sigma}d^2x\, \sqrt{g(\mathbf{x})}\; K=2 \pi \chi,
\end{equation}
where $g(\mathbf{x})$ is the determinant of the metric tensor at point $\mathbf{x}$, which in polar coordinates is simply equal to $[\sinh(\kappa r)/\kappa]^2$. The curvature being constant, Eq.~(\ref{eq_gauss_bonnet}) can be rewritten as
\begin{equation}
\label{eq_gauss_bonnet2}
A\; \kappa^2=4 \pi (g-1),
\end{equation}
which combined with Eq.~(\ref{eq_euler_poincare}) gives the deviation from $6$ of the average coordination number in terms of the density of particles $\rho=N/A$ and the curvature:
\begin{equation}
\label{eq_coordination}
c-6= \frac{3 \kappa^2}{\pi \rho}.
\end{equation}
One can apply Eqs.~\eqref{eq_euler_poincare} and \eqref{eq_coordination} to the elementary cell of the tiling of $H^2$ generated by the periodic boundary conditions (this cell corresponds to the fundamental polygon defining the periodic boundary condition, see section~\ref{sec:model}). Eq.~(\ref{eq_euler_poincare}) shows that, neglecting disclinations of charge magnitude higher than $\pi/3$~\footnote{Disclinations with such high charge magnitude have always a nonzero probability to be present at finite temperature, which is given by the Boltzmann factor $\propto e^{-\frac{|q|}{kT}}$, as the energy of a free disclination is proportional to its charge magnitude. For the system studied here, in the range of temperatures shown in Fig.~\ref{fig:defectsDensity} and for all curvatures, the total density of disclinations with charge magnitude strictly larger than $\pi/3$ is always less than $5. 10^{-3}$, whereas the density of $5$-fold and $7$-fold disclinations is roughly two orders of magnitude bigger.}, \emph{i.e.} particles with coordination number $c \leq 4$ and $c\geq 8$, the cell must contain an \emph{excess} of exactly $12(g-1)$ negative $7$-fold disclinations (the remaining $7$-fold disclinations exactly compensate the number of $5$-fold disclinations). Thus, an excess of $12$ such disclinations is present for the octagonal cell ($g=2$) and of $24$ for the $14$-gonal cell ($g=3$). This is to be contrasted with the result on a sphere where an excess of $12$ positive $5$-fold disclinations is required. In addition, Eq.~(\ref{eq_gauss_bonnet2}) shows that the area of the elementary cell depends on both the curvature and the type of boundary condition (through the genus $g$). Note then that the excess density of $7$-fold disclinations is independent of the specific fundamental polygon chosen for the periodic boundary condition and only depends on the curvature. It is indeed equal to $12(g-1)\kappa^2/[4\pi(g-1)]=3\kappa^2/\pi$.

\section{Continuum theory of defects in hyperbolic geometry}
\label{sec:continuum_theory}

In this section, we summarize the continuum theory developed in Refs.~\cite{bowick00,bowick02,travesset05,bowick06,bowick07}. We more specifically apply it to negatively curved manifolds, which is appropriate for our atomic model on $H^2$ with periodic boundary conditions. The formalism directly deals with defect degrees of freedom and starts with the elementary ones, the disclinations, from which all other topological defects can subsequently be built. Six-coordinated particles are treated using continuum elastic theory~\cite{bowick00}. At low temperature, the elastic free energy of an arbitrary disclination density on a 2D manifold $\Sigma$ of constant negative curvature $-\kappa^2$ is given by~\cite{bowick00,bowick02,bowick06}
\begin{equation}
\begin{split}
\label{eq_continuum_F_disclin}
F= &\frac{Y}{2}\int_{\Sigma} \int_{\Sigma}d^2x \sqrt{g(\mathbf{x})}\; d^2y \sqrt{g(\mathbf{y})}\; \times \\& \left(s(\mathbf{x}) + \kappa^2 \right) \frac{1}{\Delta^2}\lvert_{\mathbf{x}\mathbf{y}}\left(s(\mathbf{y}) + \kappa^2 \right) + \sum_{i=1}^{N_d}E_{core,i},
\end{split}
\end{equation}
where $Y$ is the Young modulus of the hexagonal crystal in flat space, $1/\Delta^2$ is the inverse Laplacian-squared operator, and $s(\mathbf{x})$ the disclination density, defined as
\begin{equation}
\label{eq_disclin_density}
s(\mathbf{x}) = \frac{\pi}{3\sqrt{g(\mathbf{x})}}\sum_{i=1}^{N_d}q_{i}\, \delta(\mathbf{x}- \mathbf{x}_i),
\end{equation}
where $q_i \in \mathbb{Z}^*$ (i.e., all integers except zero) is the reduced topological charge of the $i$th disclination located at $\mathbf{x}_i$ (in practice at low $T$, we only need to consider $q_i=\pm 1$) and $N_d$ is the total number of disclinations in $\Sigma$; finally, $E_{core,i}$ is a disclination core (free) energy, renormalized by thermal fluctuations~\cite{bowick00}. Note that in the present problem, Eq.~(\ref{eq_continuum_F_disclin}) is the free energy for the elementary cell of the tiling of $H^2$ associated with the periodic boundary condition. Provided interactions between cells are weak, the total free energy of the whole system is just this quantity multiplied by the (macroscopic) number of cells. This is of course different from spherical geometry where the whole system is finite. The disclination density must therefore satisfy a ``topological'' generalization of the electroneutrality constraint, namely,
\begin{equation}
\label{eq_electroneutrality}
\int_{\Sigma}d^2x \sqrt{g(\mathbf{x})}\; \left(s(\mathbf{x}) + \kappa^2 \right) = 0,
\end{equation}
even at finite temperature. By using Eq.~(\ref{eq_disclin_density}) and the Gauss-Bonnet theorem, Eq.~(\ref{eq_gauss_bonnet2}), the above expression leads to a sum rule on the total topological charge, namely
\begin{equation}
\label{eq_electroneutrality2}
\sum_{i=1}^{N_d}q_{i}= -12(g - 1).
\end{equation}
Not surprisingly, one then obtains the same result as previously derived on the basis of the Euler-Poincaré relation: there must be an excess of $12(g-1)$ disclinations with elementary charge $q=-1$ (per elementary cell) at low $T$.

The presence of the $12(g-1)$ \textit{localized} irreducible defects only imperfectly screens the curvature which is \textit{uniform} throughout space. The resulting elastic strain energy could then potentially be reduced by introducing dislocations that represent tightly bound pairs of $+1$ and $-1$ disclinations. The amplitude of the stress generated by one of the irreducible, unpaired, $-1$ (\textit{i.e.} $7$-fold) disclinations is controlled by an effective disclination charge at a geodesic distance $r$ of the disclination center. This quantity is given by
\begin{equation}
\begin{split}
\label{eq_disclin_charge}
s_{eff}(r) &= -\frac{\pi}{3} - 2 \pi \int_{0}^{r}dr' \, \frac{\sinh(\kappa r')}{\kappa}\left( - \kappa^2 \right)\\&
=-\frac{\pi}{3}\left[7 - 6 \cosh(\kappa r) \right] 
\end{split}
\end{equation}
and vanishes at the critical radius $r_c = \kappa ^{-1}\cosh^{-1}(7/6) \simeq 0.57 \kappa ^{-1}$.

It was shown in Ref.~\cite{bowick00,bowick02} for spherical geometries that there is a range of dimensionless curvatures $\kappa \sigma$ for which the strain energy associated with the stress field generated by the irreducible disclinations is minimized by the presence of strings of dislocations that radiate from each of the irreducible disclinations. This can easily be transposed to hyperbolic geometry. These strings of dislocations, whose Burgers vector is perpendicular to the string axis, terminate in the crystalline background at a distance of the order of the critical radius $r_c$. These structures have been dubbed ``grain boundary scars''~\cite{bowick00,bowick02,bausch03}. The typical number of dislocations per irreducible disclination can be estimated at zero temperature by assuming that a scar contains $m$ strings aligned along geodesics emanating from the disclination center. It has then been argued~\cite{bowick00} that dislocations along a geodesic are roughly separated by a distance $\ell(r) \simeq m \sigma/|s_{eff}(r)|$, so that the number of dislocations per irreducible disclination is given by
\begin{equation}
\begin{split}
\label{eq_disloc_number}
N_d & \simeq m \int_{0}^{r_c}dr \, \ell(r)^{-1} - cst\\&
\simeq - cst + \frac{\pi}{3}\left[7 \cosh^{-1}\left(\frac{7}{6}\right) - \sqrt{13}\right] (\kappa \sigma)^{-1}\\&
\simeq - cst + 0.3998 \, (\kappa \sigma)^{-1},
\end{split}
\end{equation}
where $cst$ denotes a constant coming from subdominant terms which depend on microscopic details (such as the Young modulus and the core energies). In contrast, the contribution proportional to $(\kappa \sigma)^{-1}$ is universal for a given geometry and underlying type of crystalline order.

The above estimate does not take into account possible interactions between grain boundary scars. In the same approximation, one can derive from Eq.~(\ref{eq_continuum_F_disclin}) the (free) energy of a boundary scar. Upon assuming that the strings radiating from an irreducible disclination and forming a given scar do not interact, the energy can be written as the sum of dislocation-dislocation pair energies, which are always repulsive, and of a disclination-dislocation energy, which for a dislocation in a geodesically aligned string at a distance $r$ from the disclination center is given by
\begin{equation}
\label{eq_disclin_disloc}
F_{Dd}(r) = \frac{Y}{12} (\kappa \sigma)^{-1}\left[\frac{\sinh(\kappa r) \ln \left[ \frac{\cosh(\kappa r) -1}{2}\right] }{1+\cosh(\kappa r)} \right],
\end{equation}
where we have assumed that the magnitude of the Burgers vector is given by $\sigma$, the lattice spacing of the underlying crystal. This expression can be obtained from that derived for spherical crystals~\cite{bowick07} by replacing $R^{-1}$ by $\mathrm{i} \kappa$. $F_{Dd}(r)$ is attractive up to a distance of order $\kappa^{-1}$. It is the competition between the energy gain coming from this attraction and the energy cost due to the repulsion between dislocations and to the dislocation core energies that results in stable grain boundary scars, at least for sufficiently small dimensionless curvature $\kappa \sigma$.

\begin{figure} 
	\begin{center}
		\includegraphics[draft=false,width=8cm]{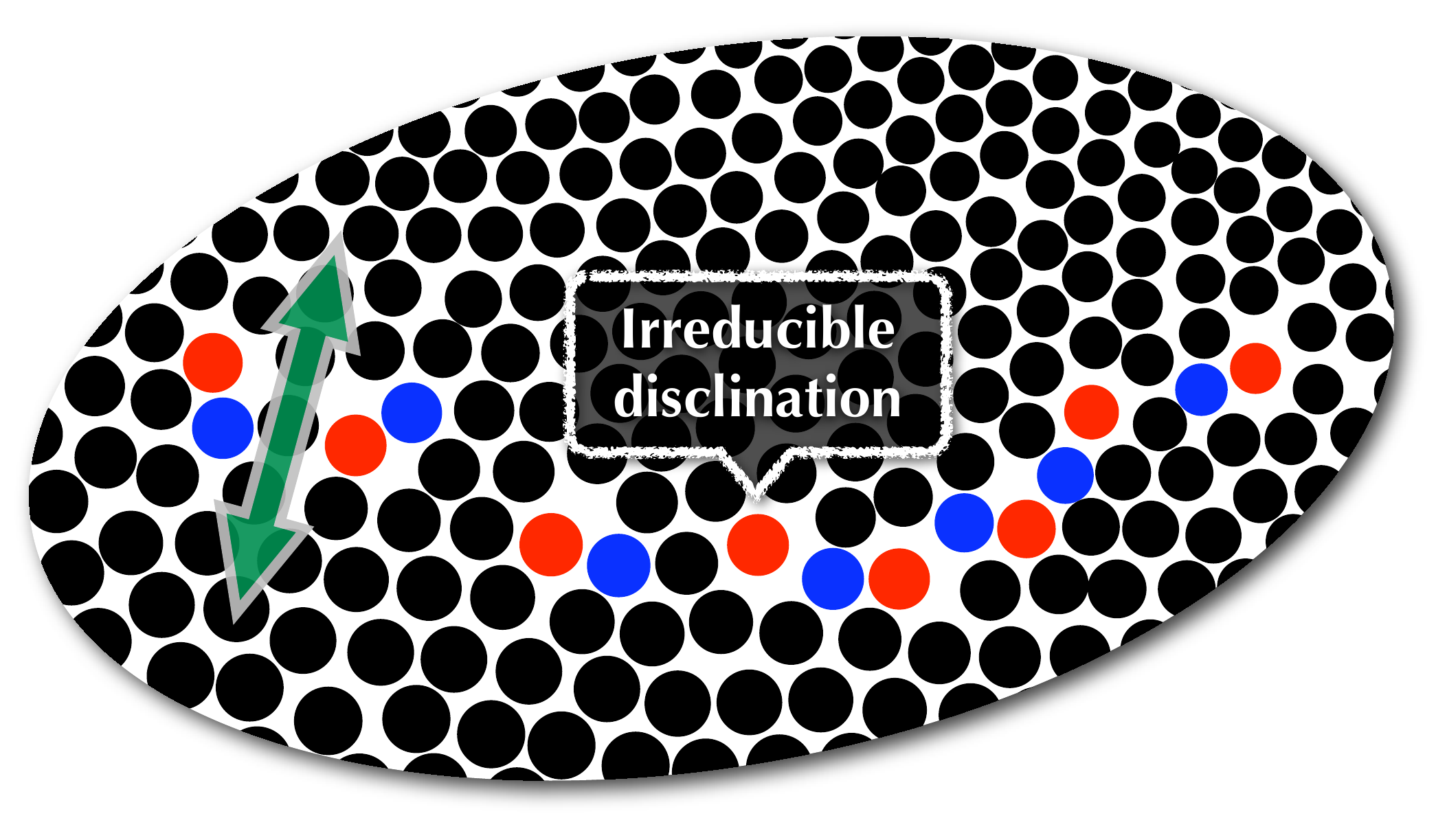}
	\end{center}
	\caption{ \label{fig:ScarSketch} (Color online) Sketch of a grain boundary scar with $2$ arms of dislocations formed by $5-7$ (blue-red) dipoles attached to a central $7$-fold (red) irreducible disclination. The glide motion takes place along the Burgers vector as indicated by the arrow.} 
\end{figure}

We now turn to the dynamics of the defects at low temperatures. We focus on the structural relaxation time $\tau$ at low but finite $T$. We again consider a situation in which there are $12(g-1)$ noninteracting grain boundary scars. A simple estimate of the structural relaxation time is provided by the time scale for the diffusion of the most loosely bound dislocation in a scar to another scar. Climb motion being energetically prohibitive, the motion of a dislocation proceeds via thermally activated glide motion (\textit{i.e.} parallel to the Burgers vector). This takes place in a potential energy that combines the Peierls potential due to the underlying lattice and the interaction potential with both the central disclination and the other dislocations of the scar~\cite{lipowsky05}. The main contribution to the activation energy barrier along the dislocation motion is due to the disclination-dislocation interaction. The latter increases as the distance between the outermost dislocation and the central disclination increases along the glide direction (which, we recall, is perpendicular to the average direction of the string: see Fig.~\ref{fig:ScarSketch}). A crude estimate for the barrier is obtained from a harmonic approximation to $F_{Dd}$ in Eq.~(\ref{eq_disclin_disloc}) for a \textit{transverse} displacement of the dislocation. For small curvature magnitude $\kappa \sigma$ and after taking into account that the outermost dislocation is originally at a distance $\sim \kappa^{-1}$ of the central disclination, this leads to
\begin{equation}
\label{eq_d_harmonic}
F_{Dd}(\kappa^{-1},y) - F_{Dd}(\kappa^{-1},0) \sim C \, Y \kappa \sigma y^2 + \cdots,
\end{equation}
where $C > 0$ is a constant of order unity and $y$ is the distance along the geodesic that is perpendicular to the string at the position of the chosen dislocation (\textit{e.g.}, along the arrow in Fig.~\ref{fig:ScarSketch}). To relax the underlying translational order, the dislocation should move a distance of order $\kappa^{-1}$ (which corresponds to the typical distance between grain boundary scars). This involves an energy barrier whose order of magnitude estimated from Eq.~(\ref{eq_d_harmonic}) is $Y\sigma \kappa^{-1}$. The final estimate for the structural relaxation time is thus
\begin{equation}
\label{eq_tau_continuum}
\tau \sim \frac{1}{D_g \kappa^2} \exp\left(\frac{E_0 + E_1 (\kappa \sigma)^{-1}}{k_B T} \right) ,
\end{equation}
where $D_g$ is a ``bare'' glide diffusion constant (\emph{i.e.} describing small displacements of a free dislocation~\cite{lipowsky05}), $E_0$ is the energy barrier due to the Peierls potential and other subdominant terms, and $E_1\propto Y \sigma^2$ corresponds to the barrier estimated above. These predictions will be compared with simulation results below.

The above arguments have been directly adapted from the theory of spherical crystals, with the necessary changes coming from the negative curvature. However, there may be a more serious difference between spherical and hyperbolic geometries due to the fact that the atomic system is of finite extent in the former and possibly of infinite extent in the latter (with periodic boundary conditions, a proper thermodynamic limit can be taken). As mentioned above, \textit{bona fide} thermodynamic phase transitions are possible in the hyperbolic plane $H^2$, but not on the sphere. Strictly speaking, freezing of the $12$ irreducible $5$-fold disclinations on a sphere in an icosahedral superlattice only occurs at zero temperature. At finite temperature, provided of course that one waits long enough for equilibrium to be achieved, the $12$ defects can move distances of the order of the sphere radius and the whole system continuously reorganizes. On $H^2$ on the other hand, one can envisage a 'crystal of defects' formed by a periodic arrangement of the irreducible $7$-fold disclinations (with or without associated grain boundary scars) on a $\{3,7\}$ lattice. The latter lattice is the only one that is compatible with a predominance of 6-fold coordination numbers. Indeed, in the hyperbolic plane, each possible lattice possesses a uniquely determined lattice spacing (for a given curvature), which is characteristic of its symmetry. Thus, there exists only one possible lattice compatible with the density of irreducible defects. Long range or quasi-long range order could then be achieved through a finite-temperature phase transition, much like the formation of an Abrikosov-like lattice of vortices in a uniformly frustrated $XY$ spin model in 2D flat space~\cite{nelson02}. Checking this possibility in simulations of our atomic liquid model is however very hard for two reasons: the difficult equilibration of the system at low temperature in the ``glassy'' liquid regime on one hand and the constraint due to the periodic boundary condition on the other. The latter constraint comes from the fact that the elementary cell of the periodic boundary condition procedure, with its $12(g-1)$ irreducible disclinations, should be compatible with the $\{3,7\}$ tiling of the hyperbolic plane with vertices centered on the irreducible defects~\footnote{By using standard hyperbolic trigonometry (see Appendix C in~\cite{sausset07}), one can compute the cell area of any $\{p,q\}$ tiling and deduce from it the associated density of vertices. A $\{3,7\}$ crystal of disclinations can only appear if the density of vertices of the $\{p,q\}$ tiling corresponding to the chosen periodic boundary condition is equal to the density of irreducible disclinations.}.

\section{Simulation results and discussion}
\label{sec:simulation_results}

\begin{figure*} 
	\begin{center}
		\captionsetup[subfloat]{margin=2.2cm,singlelinecheck=false}
		
		\subfloat[\hspace{2cm}]{\includegraphics[draft=false,scale=0.19, trim= 0cm 3.4cm 0 0]{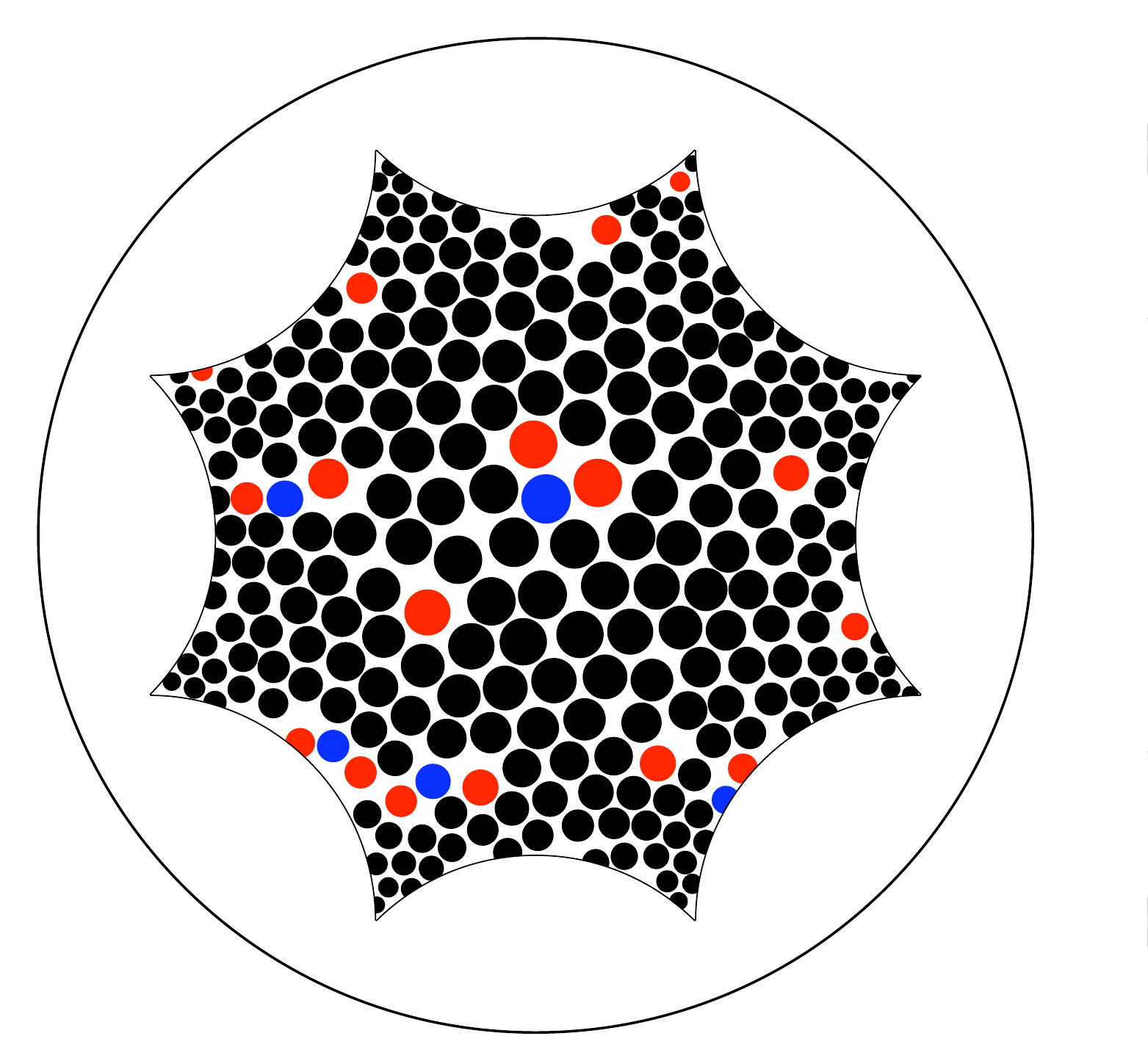}}
		
		\captionsetup[subfloat]{margin=3.5cm,singlelinecheck=false}
		
		\subfloat[\hspace{2cm}]{\includegraphics[draft=false,scale=0.38, trim= 0cm 2cm 0 0]{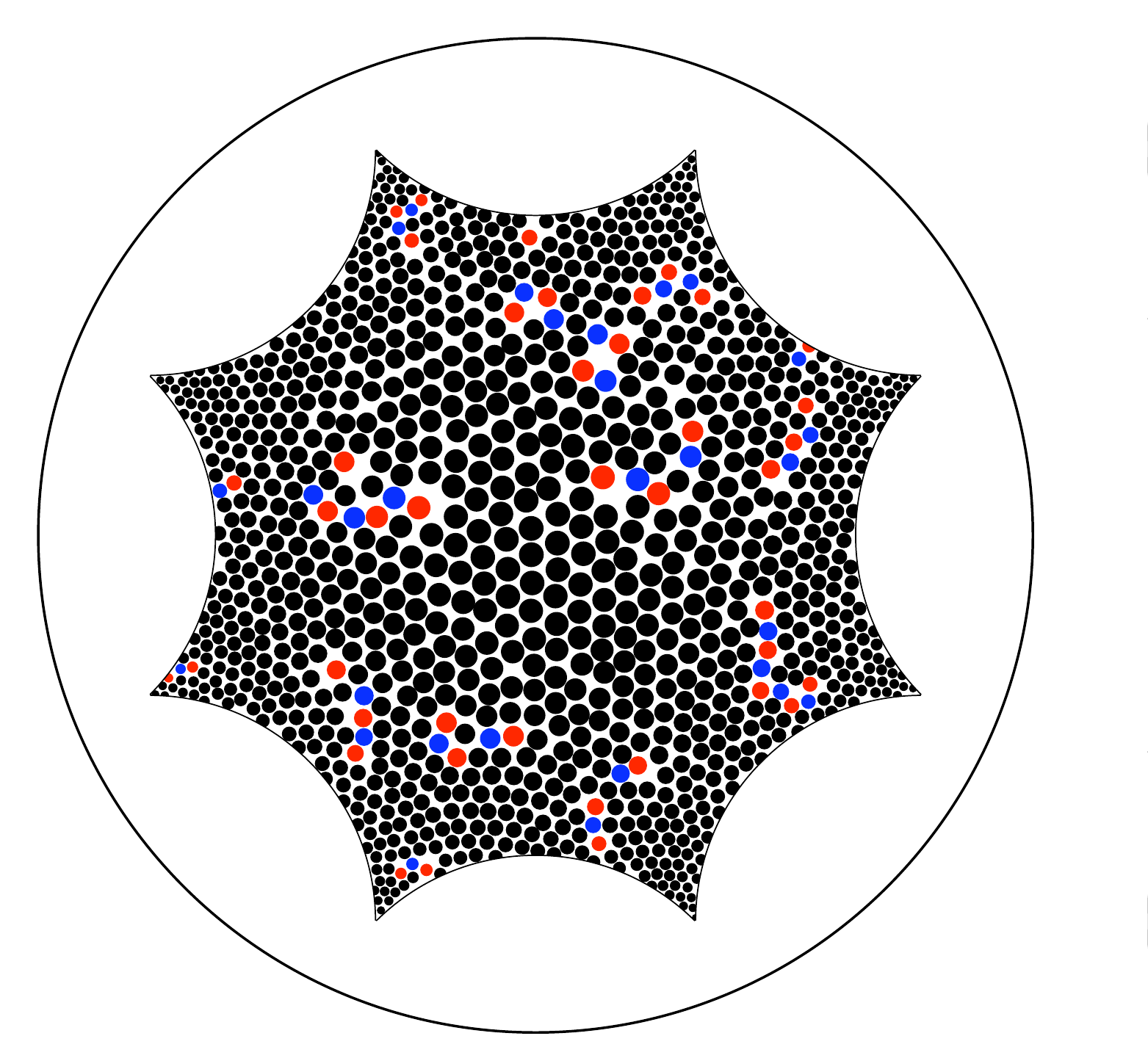}}
		\subfloat[\hspace{2cm}]{\includegraphics[draft=false,scale=0.38, trim= 0cm 2cm 0 0]{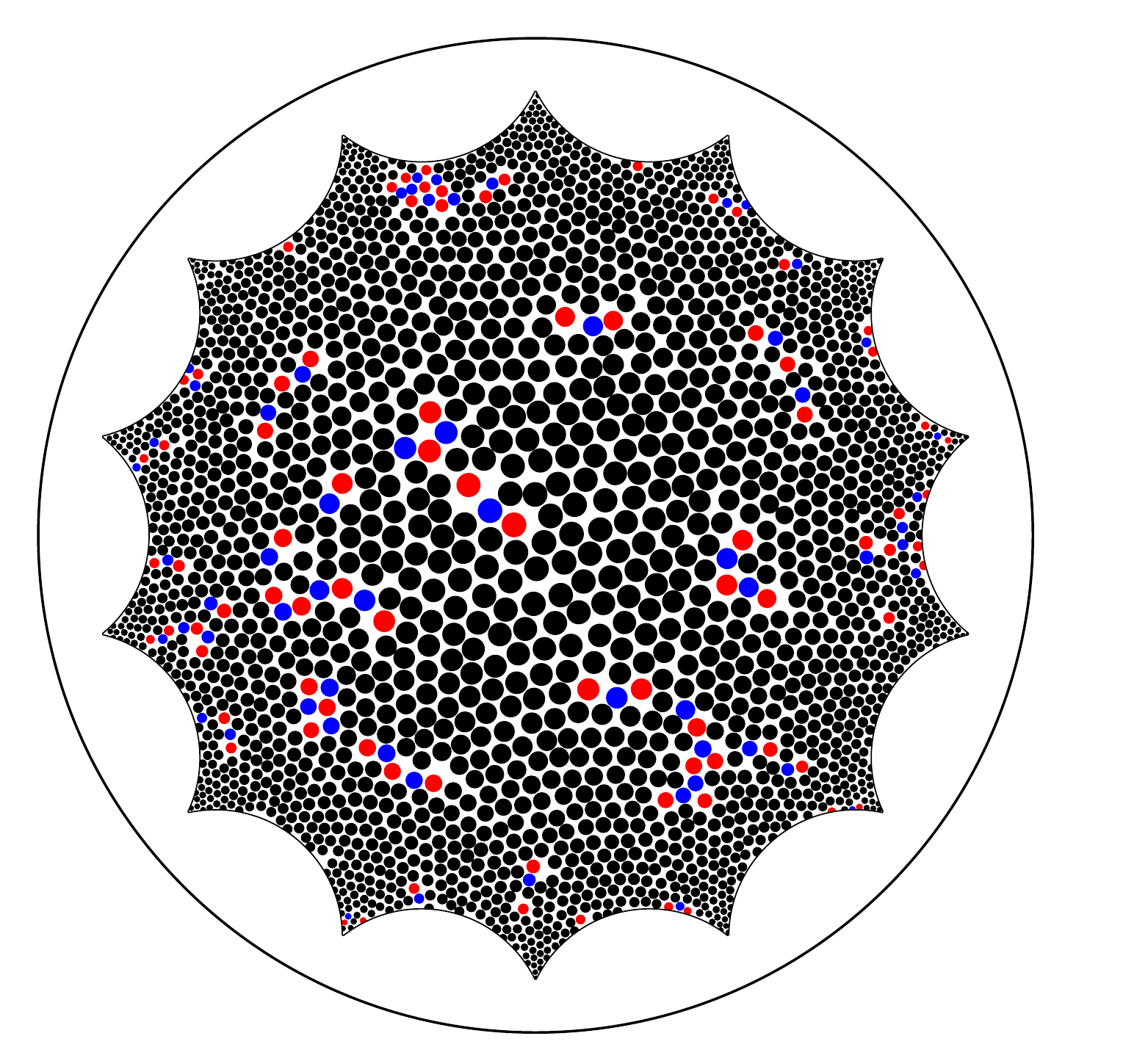}}
		
		\captionsetup[subfloat]{margin=5.7cm,singlelinecheck=false}
		
		\subfloat[\hspace{2cm}]{\includegraphics[draft=false,scale=0.76, trim= 0cm 1.1cm 0 0]{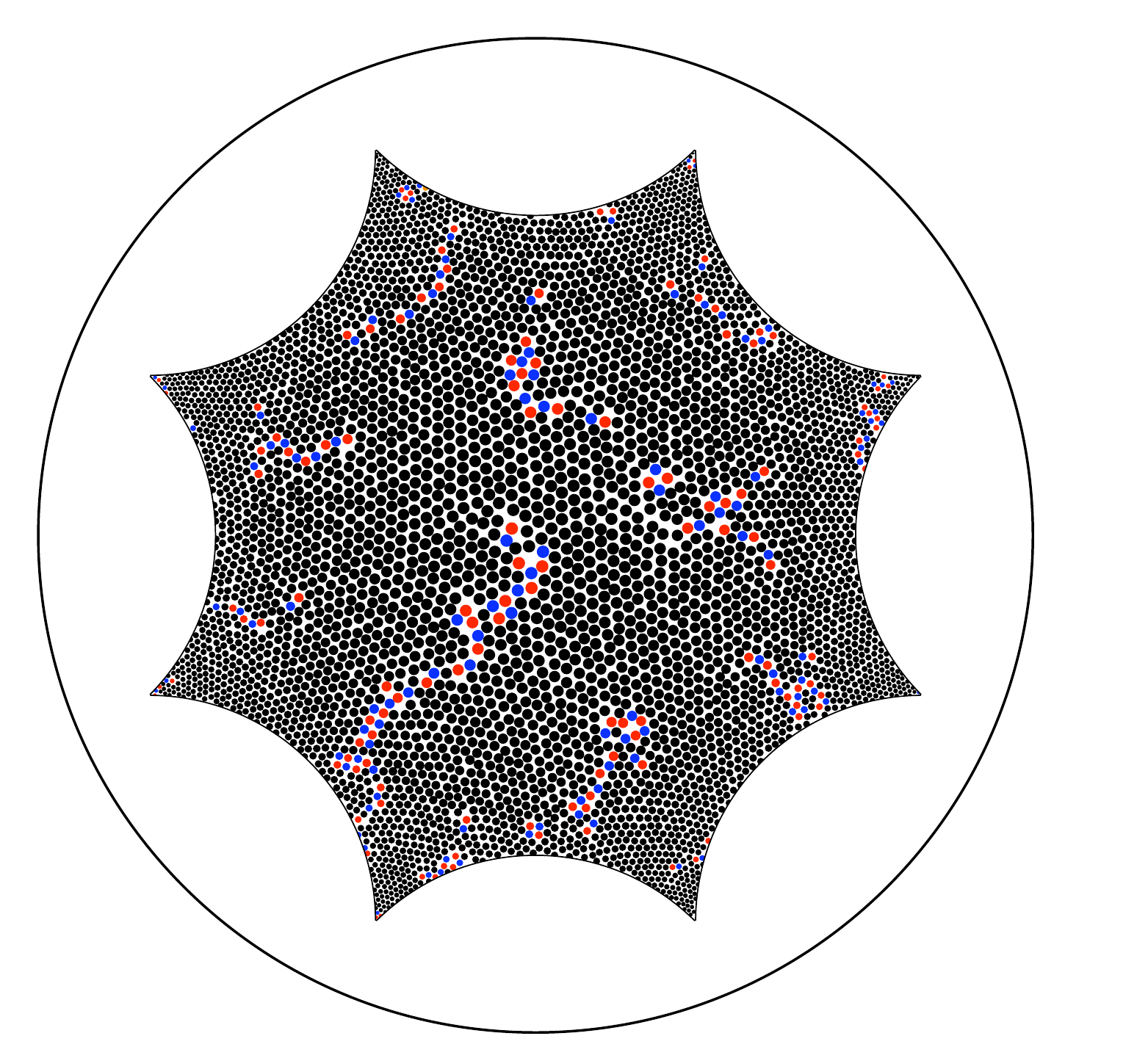}}
		
		\captionsetup[subfloat]{margin=0cm,singlelinecheck=true}
	\end{center}
	\caption{ \label{fig:lowTatomicConfs} (Color online) Low-$T$ atomic configurations showing the defect structures on $H^2$: (a) $\kappa \sigma = 0.2$, $T/T^* \simeq 0.345$, octagonal pbc; (b) $\kappa \sigma = 0.1$, $T/T^* \simeq 0.497$, octagonal pbc; (c) $\kappa \sigma = 0.1$, $T/T^* \simeq 0.520$, $14$-gonal pbc; (d) $\kappa \sigma = 0.05$, $T/T^* \simeq 0.521$, octagonal pbc. The system is shown at constant atomic size $\sigma$, which allows one to better visualize the extension of the scars as a function of curvature. Note also the difference between the two different pbc's in (b) and (c): $12$ irreducible defects are present for the octagon and $24$ for the $14$-gon.} 
\end{figure*}

As already illustrated in Fig.~1, we find that as temperature is lowered, the number of defects decreases drastically until one is left with an irreducible number of negative $7$-fold disclinations and an excess of dislocations formed by tightly bound pairs of $5$-fold and $7$-fold disclinations. We show additional low-T atomic configurations in Fig.~\ref{fig:lowTatomicConfs}, for several values of the curvature and different periodic boundary conditions (octagonal cell with $12$ irreducible negative disclinations and $14$-gonal cell with $24$ such disclinations). Note that the scars centered on the excess disclinations only become evident for small $\kappa \sigma$.

\begin{figure}
	\begin{center}
		\includegraphics[draft=false,width=8cm]{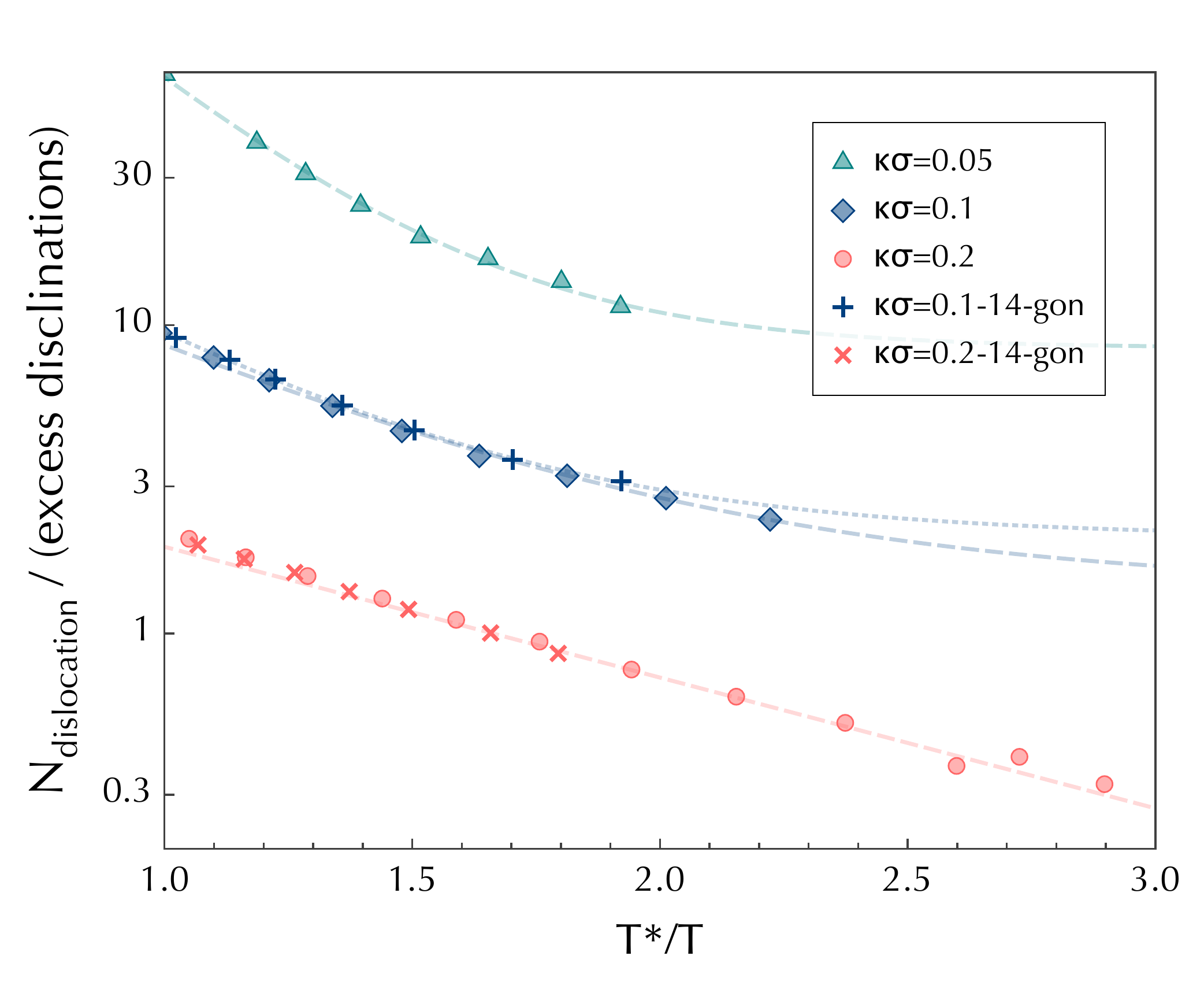}
	\end{center}
	\caption{ \label{fig:defectsDensity} (Color online) Temperature dependence of the number of $5$-fold disclinations divided by the number of irreducible $7$-fold disclinations (\textit{i.e.}, $12(g-1)$) for $T<T^*$: $\kappa \sigma = 0.2$, $\kappa \sigma = 0.1$, $\kappa \sigma = 0.05$. For $\kappa \sigma = 0.1$ and $\kappa \sigma = 0.2$, both octagonal and $14$-gonal pbc's are shown. The lines correspond to an exponential fit.} 
\end{figure}

\begin{figure}
	\begin{center}
		\includegraphics[draft=false,width=8cm]{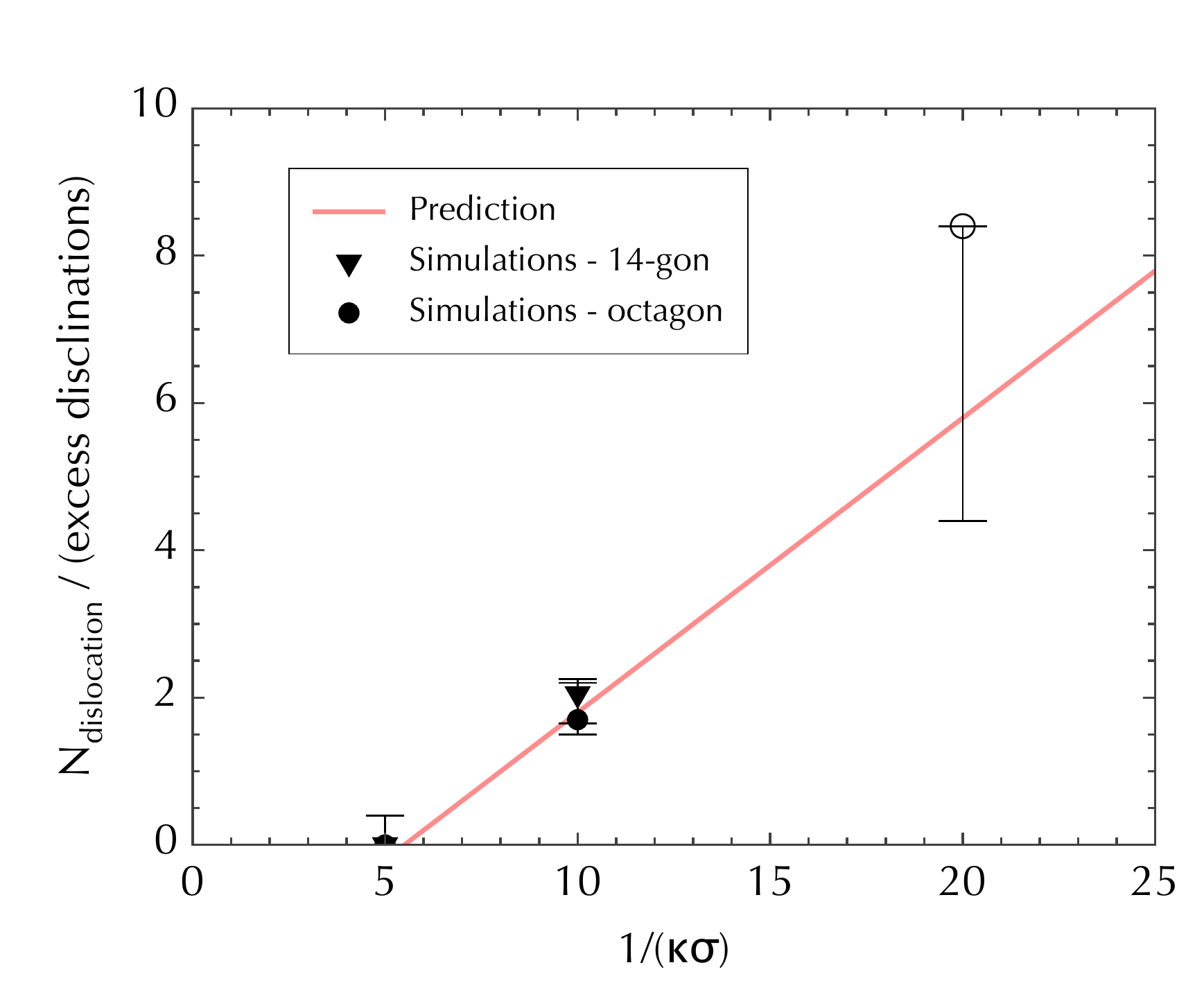}
	\end{center}
	\caption{ \label{fig:check} (Color online) Extrapolated number $N_d$ of dislocations per irreducible disclination as a function of $(\kappa \sigma)^{-1}$. The straight line is the theoretical prediction from Eq.~(\ref{eq_disloc_number}) (up to an adjustable parameter which is the intercept with the $x$-axis). The data for $\kappa \sigma = 0.05$ (octagonal pbc) is shown as an empty circle and only corresponds to an upper bound, as the liquid cannot be equilibrated at low enough $T$ (see Fig.~\ref{fig:defectsDensity}).} 
\end{figure}

To provide a more quantitative test of the theoretical prediction, we have monitored as a function of temperature the number of $5$-fold (\textit{i.e.} $q=+1$) disclinations. Below a temperature roughly equal to the ordering temperature $T^*$ in flat space~\cite{footnote}, these disclinations are all paired with $7$-fold (\textit{i.e.} $q=-1$) disclinations and, moreover, disclinations of higher charge magnitude $\vert q\vert \geq 2$ have almost completely disappeared. As seen in the plot shown in Fig.~\ref{fig:defectsDensity}, the number of 5-7 dislocations decreases with decreasing temperature and seems to saturate at a nonzero value, which we extract through an exponential fit to the data. This saturation value, divided by the number of irreducible $7$-fold disclinations in the cell (\textit{i.e.} $12(g-1)$), gives an estimate of the average number $N_d$ of dislocations which are contained at $T=0$ in a grain boundary scar radiating from an irreducible disclination. This allows us to compare the values obtained for several curvature magnitudes $\kappa \sigma$ and periodic boundary conditions with the theoretical prediction in Eq.~(\ref{eq_disloc_number}), only one adjustable parameter being needed for this exercise. The outcome is displayed in Fig.~\ref{fig:check} and shows compatibility between the simulation data and the prediction. For the smallest curvature parameter included, $\kappa \sigma = 0.05$, we are only able to provide an upper bound to $N_d$, as we cannot equilibrate the liquid at low enough temperature with the available computer resources. On the other hand, the results for the two other curvatures are robust and the comparison between the octagonal and $14$-gonal periodic boundary conditions show very little finite-size effects.

From Fig.~\ref{fig:check}, one can see that the structure in grain boundary scars no longer remains stable above a threshold curvature magnitude $\kappa_c \sigma \gtrsim 0.2$. This threshold value is nonuniversal and depends on the type of interaction potential between the particles. (In Ref.~\cite{bausch03} a similar threshold value $(a/R_c)\simeq 0.2$ has been found in an experimental system of colloidal particles on a sphere, but this may well be coincidental.) In addition, we have checked that the typical length of the arms emanating from an irreducible disclination is always of the order of, but smaller than, the theoretical cutoff radius $r_c=\kappa^{-1} \cosh^{-1}(7/6)$. Note that in the presently accessible temperature range, the system always remains in a liquid, \textit{i.e.} disordered, phase, albeit very slowly relaxing and apparently on the verge of glass formation. We have never been able to observe crystallization of the lattice of defects discussed above.

We have also investigated the low-$T$ dynamics of the system by monitoring the ``self intermediate scattering function'', $F_s(k,t)=(1/N)\sum_{j=1}^{N} \langle P_{-\frac{1}{2}+\mathrm{i}\frac{k}{\kappa}}(\cosh(\kappa d_j(t)) \rangle$, where $d_j(t)$ is the displacement of the $j$th atom during an elapsed time $t$, $P_{-\frac{1}{2}+\mathrm{i}\frac{k}{\kappa}}$ is a Legendre function of the first kind (which generalizes the standard function describing the correlation of density Fourier modes in 2D Euclidean space), and the wave-vector $k$ is chosen as usual close to $2\pi/\sigma$~\cite{sausset08,sausset09}. From $F_s(k,t)$, we have extracted the structural (translational) relaxation time $\tau$\cite{sausset08,sausset09}, which can then be studied as a function of temperature and curvature. One difficulty in trying to assess the validity of Eq.~(\ref{eq_tau_continuum}) is that the prediction only applies at low enough temperature in a regime dominated by irreducible grain boundary scars, when most thermal defects have disappeared and growth of the hexatic/hexagonal order has saturated. This occurs below a crossover temperature $T_X$, which in principle may be curvature-dependent. For the limited range over which we have simulation data (mostly, $\kappa \sigma = 0.1$ and $0.2$), $T_X$ appears to vary weakly with curvature and we replace Eq.~(\ref{eq_tau_continuum}) by the following expression for $T\lesssim T_X$,
\begin{equation}
\label{eq_tau_simu}
\tau \simeq \tau_X \exp\left(\left[ E_0 + E_1 (\kappa \sigma)^{-1}\right] \left[\frac{1}{k_B T}- \frac{1}{k_B T_X} \right] \right) ,
\end{equation}
where $\tau_X \equiv \tau(T_X)$ depends on curvature. The latter dependence can be tentatively estimated by assuming that the relaxation at $T_X$ is governed by the \textit{bare} diffusion of dislocations to a distance of order $\kappa^{-1}$,\textit{i.e.} $\tau_X \simeq D_g^{-1} \kappa^{-2}$.

\begin{figure}
	\begin{center}
		\includegraphics[draft=false,width=8cm]{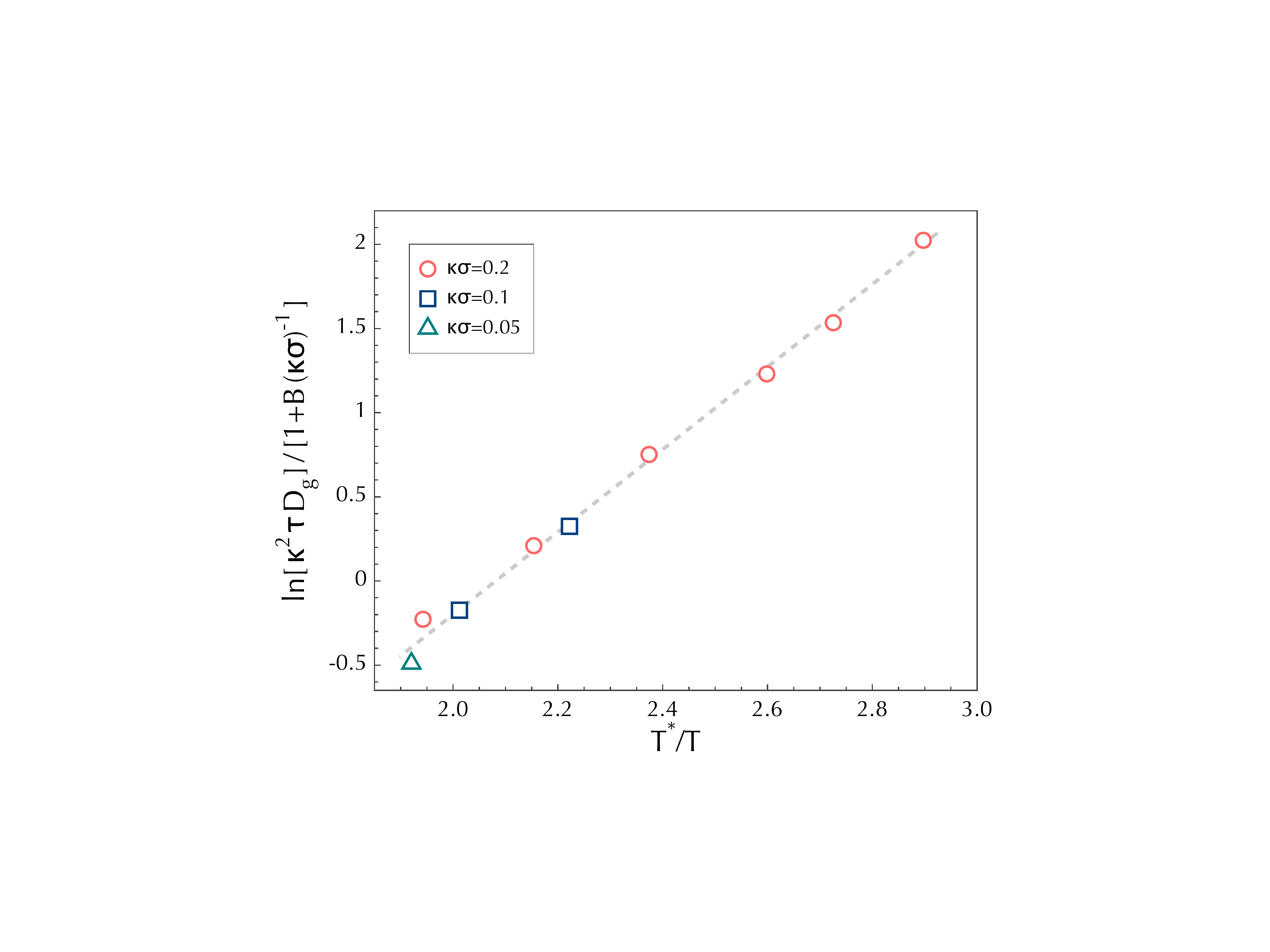}
	\end{center}
	\caption{ \label{fig:relaxationTime} (Color online) Scaled plot of the translational relaxation time $\tau(\kappa \sigma,T)$ in the low-$T$ regime near and below an estimated crossover temperature $T_X/T^* \simeq 0.5$: $\ln[\kappa^2\tau D_g]/[1+B (\kappa \sigma)^{-1}]$ versus $T^*/T$, where $B\simeq 0.1$ and $D_g\simeq 0.76 \, \sigma^2 / \tau_0$ are determined for providing the best collapse to a straight line. ($\tau_0$ is the microscopic timescale associated with the interaction potential.) The plot is compatible with Eq.~(\ref{eq_tau_simu}).} 
\end{figure}

In Fig.~\ref{fig:relaxationTime}, we test Eq.~(\ref{eq_tau_simu}) by plotting $\ln(\tau D_g \kappa^2)/[1+B(\kappa \sigma)^{-1}]$ versus $1/T$ for $T$ below a $T_X$ estimated around $0.5 T^*$ and with the constants $B$ and $D_g$ adjusted to provide the best straight line. Notwithstanding several adjustable parameters and a limited domain of available data, the dependence of the structural relaxation is at least compatible with the above prediction. However, a more extensive data set would clearly be necessary to provide a more rigorous test of this finding.

\section{Conclusion}
\label{sec:conclusion}

In this article we have studied the low-temperature structure and dynamics of a Lennard-Jones liquid model embedded in manifolds of constant negative curvature and we have compared computer simulation results with predictions of a continuum theory of defects in curved space. In the regime where the spatial extension of the local hexagonal order has saturated due to the frustration effect, we find that the atomic configurations consist of isolated defect structures formed by an irreducible finite density of curvature-induced, negative disclination from which emerge finite-length strings of dislocations. Such ``grain boundary scars'' are unique to curved geometries. The existence and the properties of these scars are in agreement with the theoretical predictions, which so far had only been confirmed for positively curved (spherical) geometry. Questions remain open for future work, such as an improved description of the long-time relaxation associated with large-distance diffusion of the defects and the possible occurrence of crystals of defects and scars.

\begin{acknowledgements}
	Work by DRN was supported by the National Science Foundation, through grant DMR-0654191 and via the Harvard Materials Research Science and Engineering Center through grant DMR-0820484.
\end{acknowledgements}

\end{document}